\documentstyle[epsf]{mn}
\def\gtrsim{\mathrel{\hbox{\rlap{\hbox{\lower4pt\hbox{$\sim$}}}\hbox{$>$}}}}
\input epsf.sty
\title[Radio Observations of J1324$-$3138]{Radio Properties
of the Shapley Concentration. \\
II. J1324$-$3138: a remnant of a radio galaxy \\
in the Abell cluster A3556?}

\author[T. Venturi et al.]{
T. Venturi$^1$, 
S. Bardelli$^2$, 
R. Morganti$^1$, 
R.W. Hunstead$^3$ \\
$^1$ Istituto di Radioastronomia, CNR,
via Gobetti 101, I-40129 Bologna, Italy\\
$^2$ Osservatorio Astronomico di Trieste, 
Via G.B. Tiepolo 11, I34131-Trieste, Italy \\
$^3$ School of Physics, University of Sydney, NSW 2006, Australia \\
E-mail: tventuri@astbo1.bo.cnr.it
}
\date{Received XX; Accepted XX}
\begin{document}

\maketitle

\begin{abstract}  

In this paper we present a detailed study of the radio galaxy J1324$-$3138,
located at a projected distance of 2$^{\prime}$ from the centre of the
Abell cluster of galaxies A3556, belonging to the core of the
Shapley Concentration, at an average redshift z=0.05.
We have observed J1324$-$3138 over a wide range of
frequencies: at 327 MHz (VLA), at 843 MHz (MOST), and at 1376 MHz, 2382 MHz,
4790 MHz and 8640 MHz (ATCA).

Our analysis suggests that J1324$-$3138 is a remnant of a tailed radio galaxy,
in which the nuclear engine has switched off and the radio source is now at a 
late stage of its evolution, confined by the intracluster gas.
The radio galaxy is not in pressure equilibrium with the external medium,
as it is often found for extended radio sources in clusters of galaxies.
We favour the hypothesis that the lack of observed polarised radio emission 
in the source is due to Faraday rotation by a foreground screen, i.e. 
the source is seen through a dense cluster gas, characterised
by a random magnetic field. 

An implication of the head-tail nature of the source is 
that J1324$-$3138 is moving away from the core of
A3556 and that possibly a major merging event between the core of A3556 and the
subgroup hosting J1324$-$3138 has already taken place.

\end{abstract}


\section{ Extended radio sources in clusters of galaxies and 
J1324$-$3138}

Extended radio emission associated with galaxies in clusters is
often characterised by morphologies which reflect the interaction between
the radio emitting plasma and the local environment in the cluster.
Head-tail sources are usually associated with non-dominant cluster
galaxies moving at a considerable speed within the cluster. Their 
morphologies are then explained as the result of ram 
pressure exerted by the intergalactic medium on the double sided radio 
emission (see for example O'Dea \& Owen 1985a and 1985b, and
Owen 1996, for a recent review). Wide-angle tail radio galaxies,
on the other hand, are more difficult to account for with the above mentioned
model, since they are usually associated with massive and dominant
cluster galaxies, with much lower peculiar velocities with respect to the
cluster mean. 
Beyond ram pressure, it is now accepted that large flows of hot gas 
could provide a wind within clusters of galaxies able to bend straight
jets into wide-angle tail morphologies (Owen 1996 and references therein).

\noindent
The study of extended galaxies in clusters is important for a variety of 
reasons. The morphology and the direction of the extension may
give important information on the dynamics of the galaxy, such as, for
example, the direction of the motion projected on the plane of the sky.
Furthermore, the non-thermal pressure in the tails of radio emission can
be compared to the thermal pressure exerted by the intracluster gas,
in those cases where X-ray data are available to provide estimates of the
temperature and pressure. This is crucial for studying the interaction
between the radio emission and the external gas, and for deriving information
on the evolution of radio sources in clusters of galaxies, as well as
the influence of the cluster dynamics (such as, for example, merging processes) 
on the radio properties of the cluster. Last but not least, 
the observed 
polarisation properties of the radio emission may give information on the
intracluster magnetic field and its structure.

\medskip
In this paper we present a detailed study of the extended radio 
galaxy J1324$-$3138 (RA$_{J2000} = 13^h24^m01^s$, 
DEC$_{J2000} = -31^{\circ}38^{\prime}$),
located in the central region of the Abell cluster A3556.
It was first observed  in a radio survey of the clusters of galaxies in the 
Shapley Concentration core
carried out at 843 MHz with the Molonglo Observatory Synthesis 
Telescope (MOST) and at 1376 MHz with the Australia Telescope
Compact Array (ATCA) (Venturi et al. 1997, hereinafter Paper I).
This work is part of a larger project whose aim is to study the radio/optical
properties of the clusters in the core of the Shapley Concentration,
in particular the chain formed by A3556-A3558-A3562 (Venturi et al. 1998),
both from a statistical point of view
and through a detailed analysis of the physical properties of the 
extended radio galaxies in these clusters.

In Figure 1 the superposition of the radio isophotes on the Digitised Sky 
Survey shows that the radio component located at the south west extremity 
of the extended radio emission is
coincident with the nucleus of the 15.6 magnitude galaxy \#5975 in the COSMOS
catalogue (RA$_{J2000}$ = $13^h23^m57.5^s$, 
DEC$_{J2000}$ = $-31^{\circ}38^{\prime}45^{\prime\prime}$).
Its radial velocity 
velocity v = 15054 km s$^{-1}$ (Stein 1996) establishes that it belongs to
A3556 ($<v> = 14357$ km s$^{-1}$, Bardelli et al. 1998).
This, coupled with the fact that only a few very faint optical objects fall
within the envelope of the radio emission,
led us to the conclusion that J1324$-$3138 is an extended,
possibly head-tail, radio galaxy located in the vicinity of the cluster centre
(see Paper I).

In Section 2 we present the observational data.
In Section 3 the morphology of the source is
described and analysed, and in Section 4 a detailed study of the synchrotron
spectrum is carried out. The nature of the source, its
relation to the cluster of galaxies A3556 
and its implications
for cluster merging and formation is discussed in Section 5.

Throughout the paper 
we use a Hubble constant of H$_0$ = 100 km s$^{-1}$Mpc$^{-1}$.
At the redshift of the cluster this implies that 1$^{\prime\prime}$ = 0.67 kpc.
%
% FIGURE 1. 
\begin{figure}
\epsfysize=8.5cm
\epsfbox{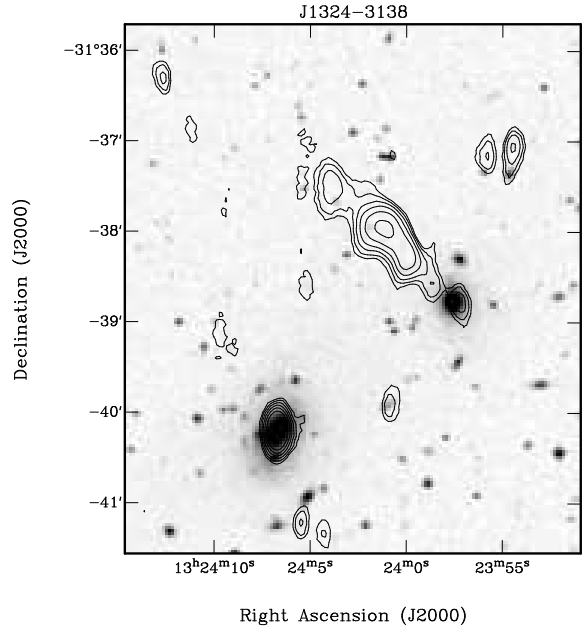}
\caption[]{
4790 MHZ radio isophotes of the extended radio galaxy 
J1324$-$3138 and of the nearby radio galaxy J1324$-$3140, associated with the 
dominant cD galaxy in the centre of A3556 (see Paper I), superimposed on the 
DSS optical image. The resolution of the image is 
$20^{\prime\prime} \times 10^{\prime\prime}$, p.a. $0^{\circ}$.}
\end{figure}
\vskip 1.0truecm
\noindent
\section{Radio observations, data reduction and images}

%
%
%
%---- TABLE 1
%
\begin{table*}
\caption[]{ Journal of the Observations }
\begin{flushleft}
\begin{tabular}{crcccrc}
\hline\noalign{\smallskip}
   Date & Frequency & Array &  Configuration & u-v range & Time & HPBW   \\
        &     MHz   &       &                &    m      &  hr  & 
$^{\prime\prime}$ \\
\noalign{\smallskip}
\hline\noalign{\smallskip}
    26-Jan-96 &  327  &  VLA   & CnB  &  46 - 2760 & 3   & $59 \times 45$  \\
    24-Mar-92 &  843  &  MOST  &      &  15 - 1571 & 12  & $82\times 43$   \\
    15-May-92 &  843  &  MOST  &      &  15 - 1571 & 12  & $82\times 43$   \\
    24-Sep-94 & 1376  &  ATCA  & 1.5D & 107 - 1439 & 12  & $16\times 40$   \\
    02-Oct-94 & 1376  &  ATCA  &  6C  & 153 - 6000 & 12  & $5 \times 10$   \\
    03-Jan-95 & 1376  &  ATCA  &  6A  & 337 - 5939 & 12  & $6.5\times9.8$  \\
    03-Jan-95 & 2382  &  ATCA  &  6A  & 337 - 5939 & 12  & $3.4\times5.3$  \\
    01-Mar-96 & 4790  &  ATCA  & 1.5B &  31 - 1286 & 12  & $13.1\times4.9$ \\
    22-Apr-96 & 4790  &  ATCA  & 375  &  31 - 459  & 12  & $36.7\times17.0$\\
    01-Mar-96 & 8640  &  ATCA  & 1.5B &  31 - 1286 & 12  & $7.1\times2.7$  \\
    22-Apr-96 & 8640  &  ATCA  & 375  &  31 - 459  & 12  & $20.3\times9.6$ \\
\noalign{\smallskip}
\hline
\end{tabular}
\end{flushleft}
\end{table*}

We have observed J1324$-$3138 over a wide range of radio frequencies, 
from 327 MHz (91.7 cm) to 8640 MHz (3.5 cm) with various arrays. The logs of 
the observations and the arrays used are summarised in Table 1. 
All flux density values given in the figure captions and in Table 2 are 
corrected for the primary beam attenuation and were computed by means of
the task TVSTAT in AIPS. Uncertainties in the flux density measurements 
are of the order of few percent at all frequencies.

\medskip
\noindent
\subsection{VLA observations at 327 MHz}

Observations at 327 MHz were carried out with the VLA in the hybrid CnB 
configuration.  The field of view of the present observations has a radius of
$\sim 2^{\circ}$, centred on J1324$-$3138 itself.
J1316$-$336 and 3C286 were used as phase and amplitude
VLA calibrators respectively. The data reduction was carried out using 
the standard procedure (calibration, Fourier inversion, clean and restore)
of the AIPS package. Contour plots of the final image are given in Figure
2 and details of the image are given in the figure caption.
%
% FIGURE 1. 
\begin{figure}
\epsfysize=8.5cm
\epsfbox{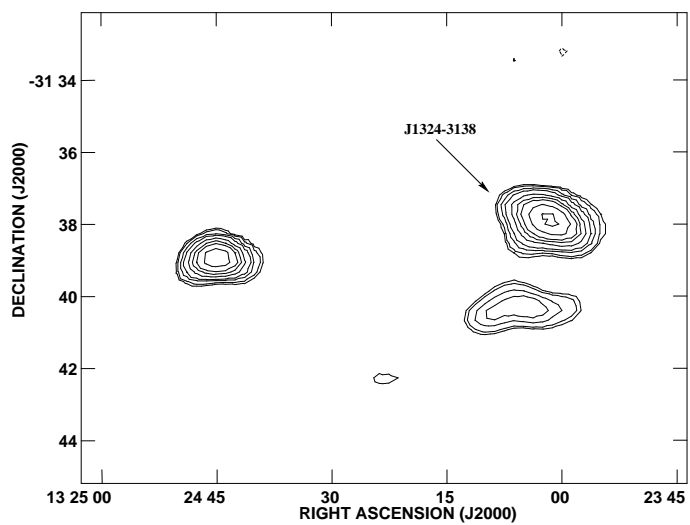}
\caption[]{
327 MHz VLA (CnB configuration) contour map of the
central region of A3556. The resolution
is $59.4^{\prime\prime} \times 45.2^{\prime\prime}$, p.a. $80.5^{\circ}$.
The total flux density in the map is 230 mJy,
and the noise is 1.9 mJy/beam.
Contour levels are -8, 8, 10, 15, 20, 30, 40, 50, 70, 100 mJy/beam.}
\end{figure}

\subsection{MOST observations at 843 MHz}

For details concerning these observations and data reduction 
we refer to Paper I, where these data were first 
presented. In Table 1 the log of the observations is reported. 
Contour plots of the MOST image are given in Figure 3 and details of
the image are given in the figure caption.
%
%
% FIGURE 3.
\begin{figure}
\epsfysize=8.5cm
\epsfbox{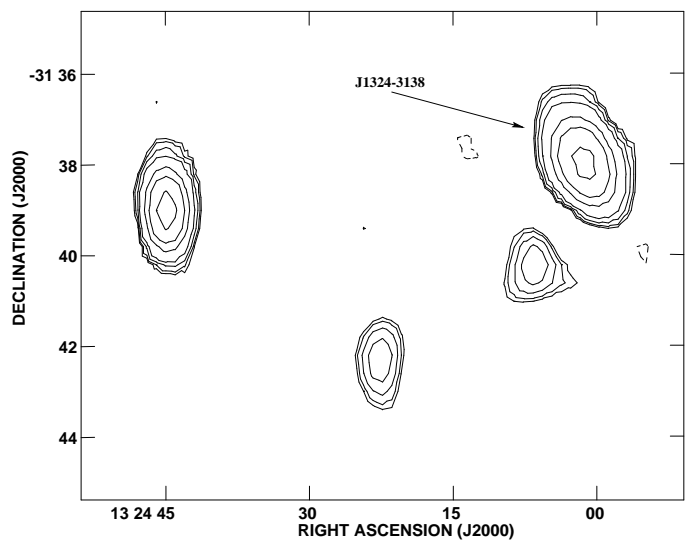}
\caption[]{
843 MHz MOST contour image of the same region as in Figure 3. 
The resolution is  
$82^{\prime\prime} \times 43^{\prime\prime}$, in p.a. $0^{\circ}$.
Contour levels are -4, 4, 5, 7, 10, 20, 30, 50 mJy/beam. 
The total flux density of J1324$-$3138
is 80 mJy. The rms noise is 1.4 mJy/beam.}
\end{figure}
\subsection{ATCA observations at 1376 MHz and 2382 MHz}

J1324$-$3138 was observed at these two frequencies as part of a larger project,
whose aim was to image and study the whole A3556 region (Paper I).
The source was observed with three array configurations,
1.5D, 6A and 6C, which differ in the minimum and maximum baseline length 
(see Table 1). The data were combined after calibration in order to increase 
the sensitivity.
We refer to Paper I for details concerning the
observing  strategy and data reduction.
Here we present images of J1324$-$3138 at 1376 MHz at increasing resolution
(Figures 4a and 4b) and at 2382 MHz (Figure 5). Details of the images 
are given in the figure captions.
%
%
% FIGURE 4. 
\begin{figure}
\epsfysize=8.5cm
\epsfbox{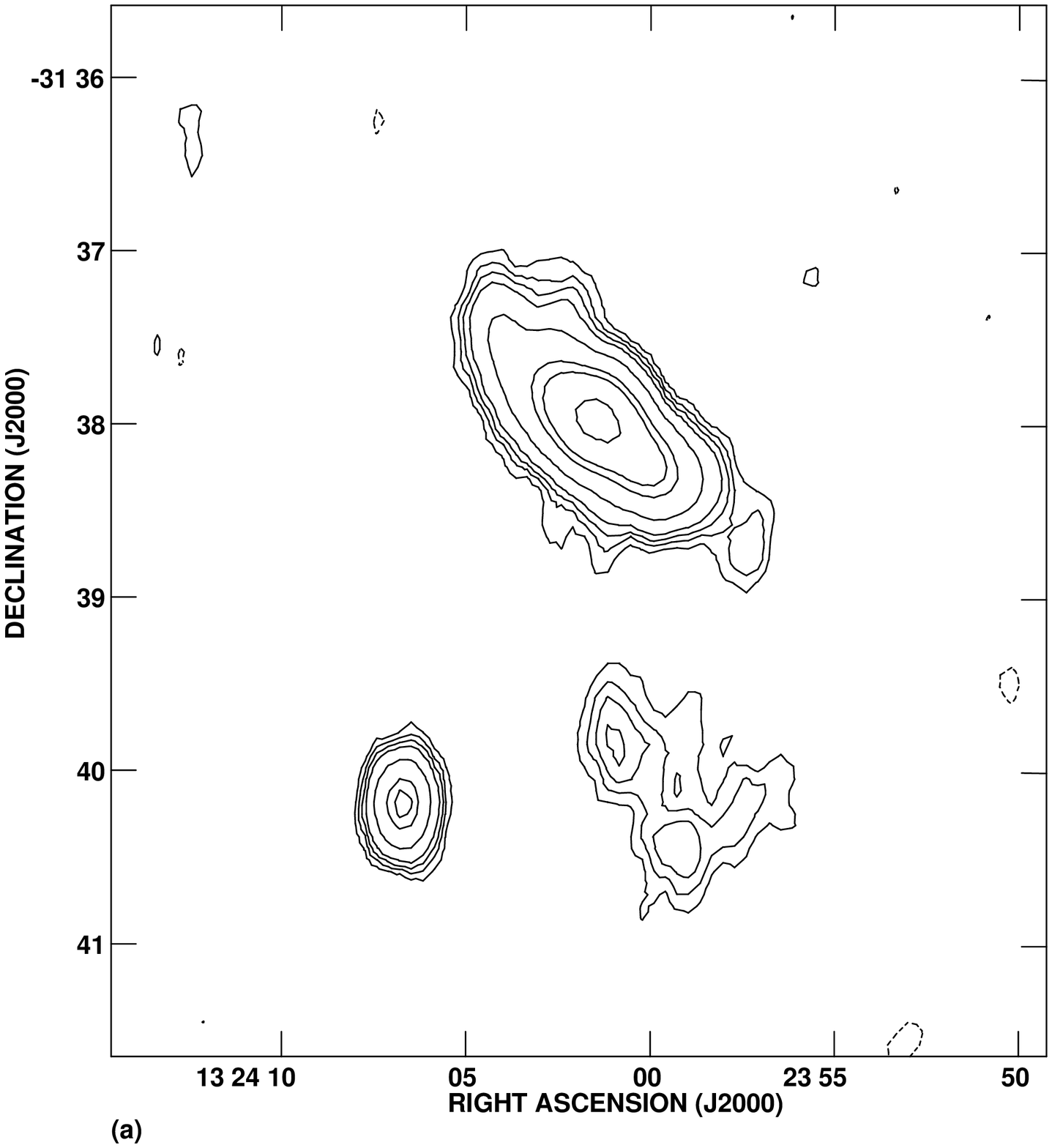}
\epsfysize=8.5cm
\epsfbox{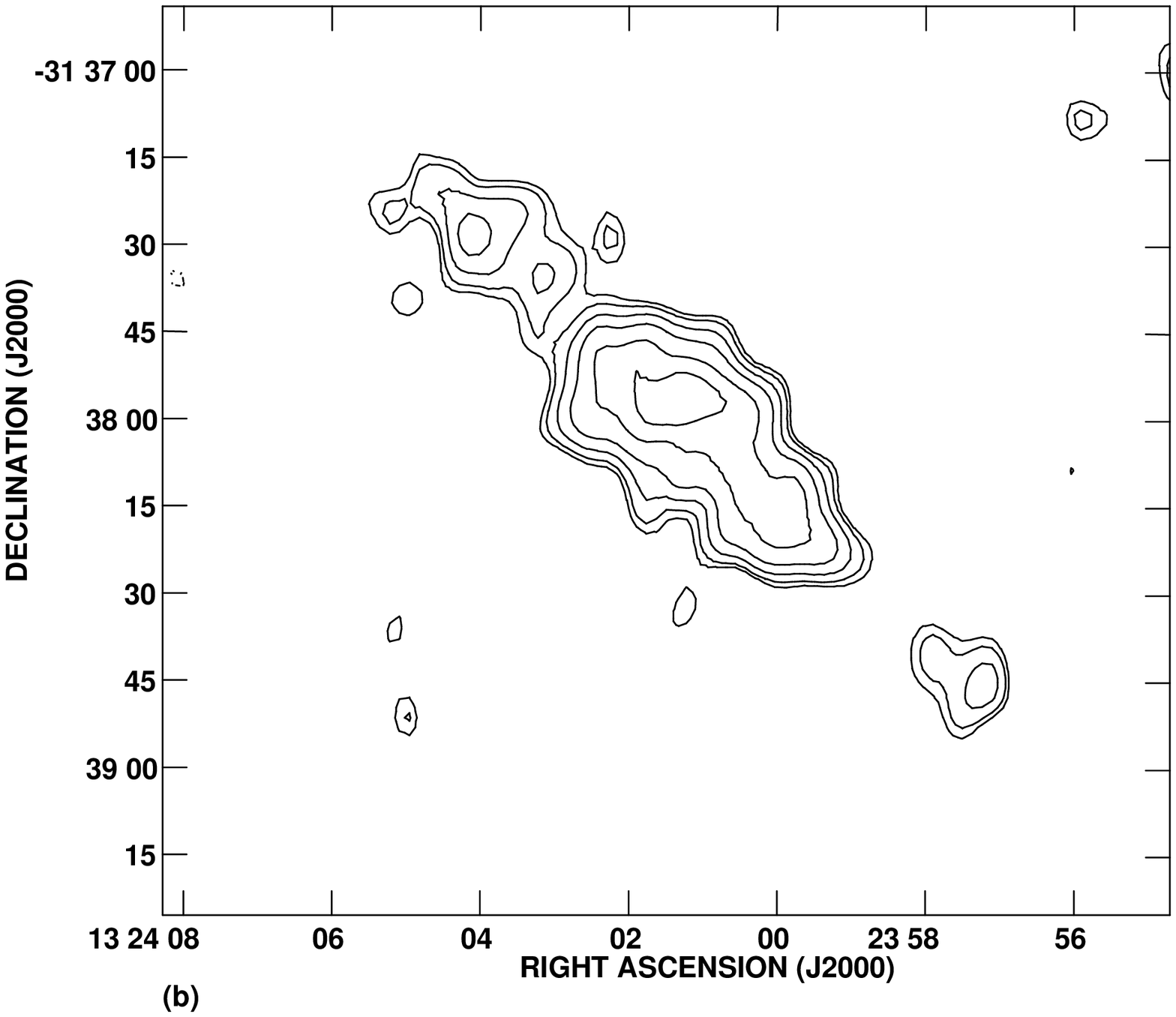}
\caption[]{
(a) Natural weighted 1376 MHz ATCA image obtained combining by data 
from the 6C and 1.5D arrays. 
The resolution is $25.8^{\prime\prime} \times 15.6^{\prime\prime}$, 
p.a. $-1.2^{\circ}$. 
The total flux density is 37 mJy.
Contour levels are -0.3, 0.3, 0.5, 0.7, 1, 2, 4, 5, 8 mJy/beam.
The rms noise in the image is 0.11 mJy/beam.
\par\noindent
(b) Full resolution 1376 MHz ATCA image, obtained by combining data 
from all arrays (6C, 6A and 1.5D). 
The resolution is $10.2^{\prime\prime} \times 5.9^{\prime\prime}$, 
p.a. $0.26^{\circ}$.
The total flux density 
is 41.2 mJy. The rms noise in the image is 0.14 mJy/beam.
Contour levels are -0.4, 0.4, 0.5, 0.7, 1, 1.5, 2 mJy/beam.}
\end{figure}
%
%
%
% FIGURE 5.
\begin{figure}
\epsfysize=8.5cm
\epsfbox{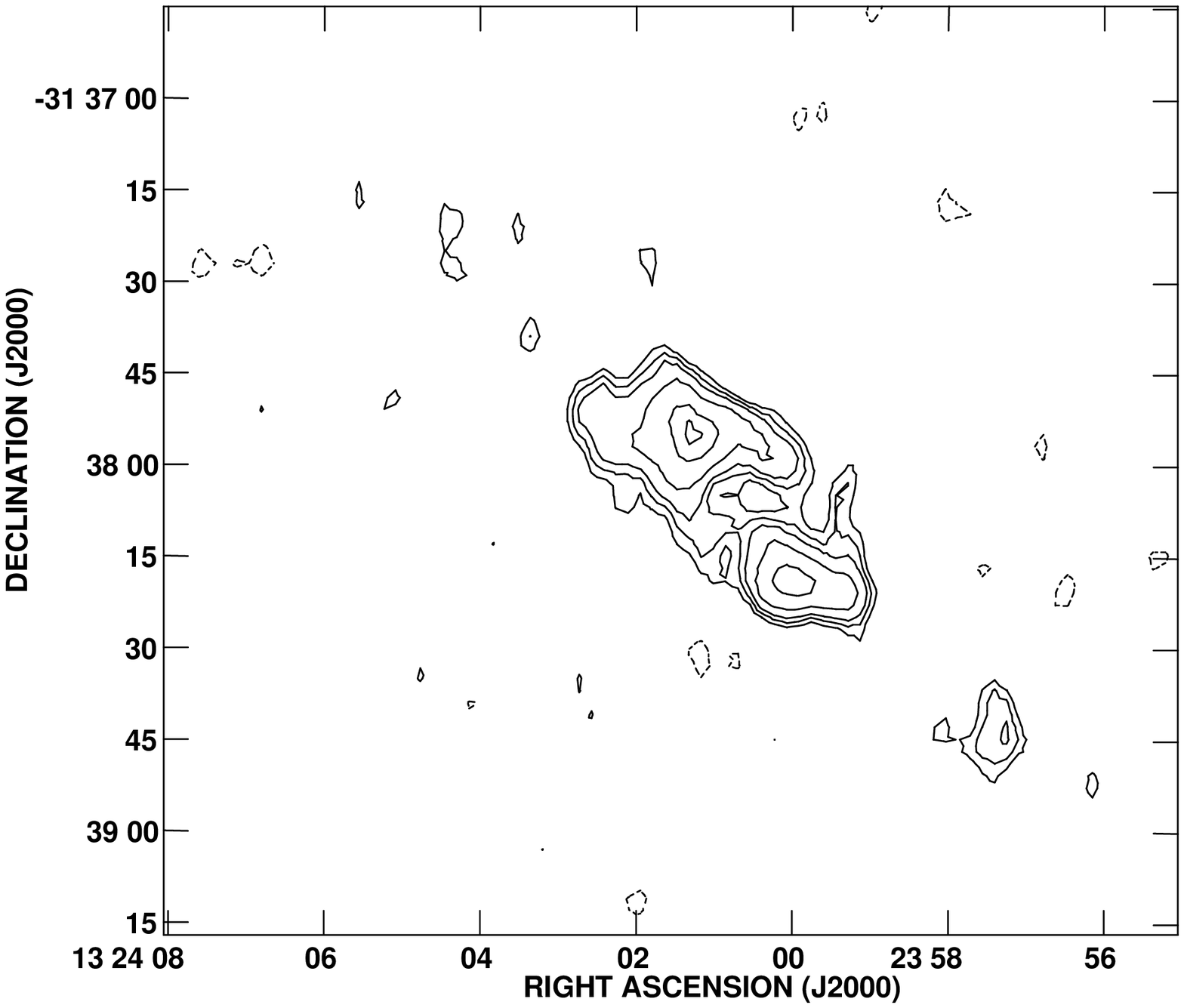}
\caption[]{
Natural weighted 2382 MHz ATCA image.
The resolution is $9.8^{\prime\prime} \times 6.5^{\prime\prime}$, 
p.a. $0^{\circ}$.
The total flux
density in the map is 13.5 mJy. Contour levels
are -0.3, 0.3, 0.4, 0.5, 0.7, 1, 1.15 mJy/beam.
The rms noise in the image is 0.16 mJy/beam.}
\end{figure}

\subsection{ATCA observations at 4790 MHz and 8640 MHz}

We observed J1324$-$3138 with the Australia Telescope Compact Array (ATCA)
simultaneously at 4790 MHz and 8640 MHz for 12 hours using the two
array configurations 1.5B and 375m (see Table 1), with an observing 
bandwidth of 
128 MHz at each frequency. The data were reduced by means of the MIRIAD
software (Sault, Teuben \& Wright 1995), which minimises 
bandwidth smearing effects, since it allows a large
number of channels to be handled separately. Data taken
with the two different array configurations were analysed independently at 
each frequency, and were subsequently combined in order to increase the
sensitivity. 

\noindent
We used B1327$-$311 as amplitude calibrator at both frequencies. Its assumed
flux is 0.63 Jy and 0.56 Jy respectively at 4790 MHz and 8640 MHz.

\noindent
The final combined datasets were then transferred to the AIPS
package and the final images were obtained by means of the task IMAGR.
The images obtained with natural and uniform weighting are given in 
Figures 6a-b and
7a-b at 4790 MHz and 8640 MHz respectively.
Details of the images are given in the figure captions.
%
%
%
% FIGURE 6. 
\begin{figure}
\epsfysize=8.5cm
\epsfbox{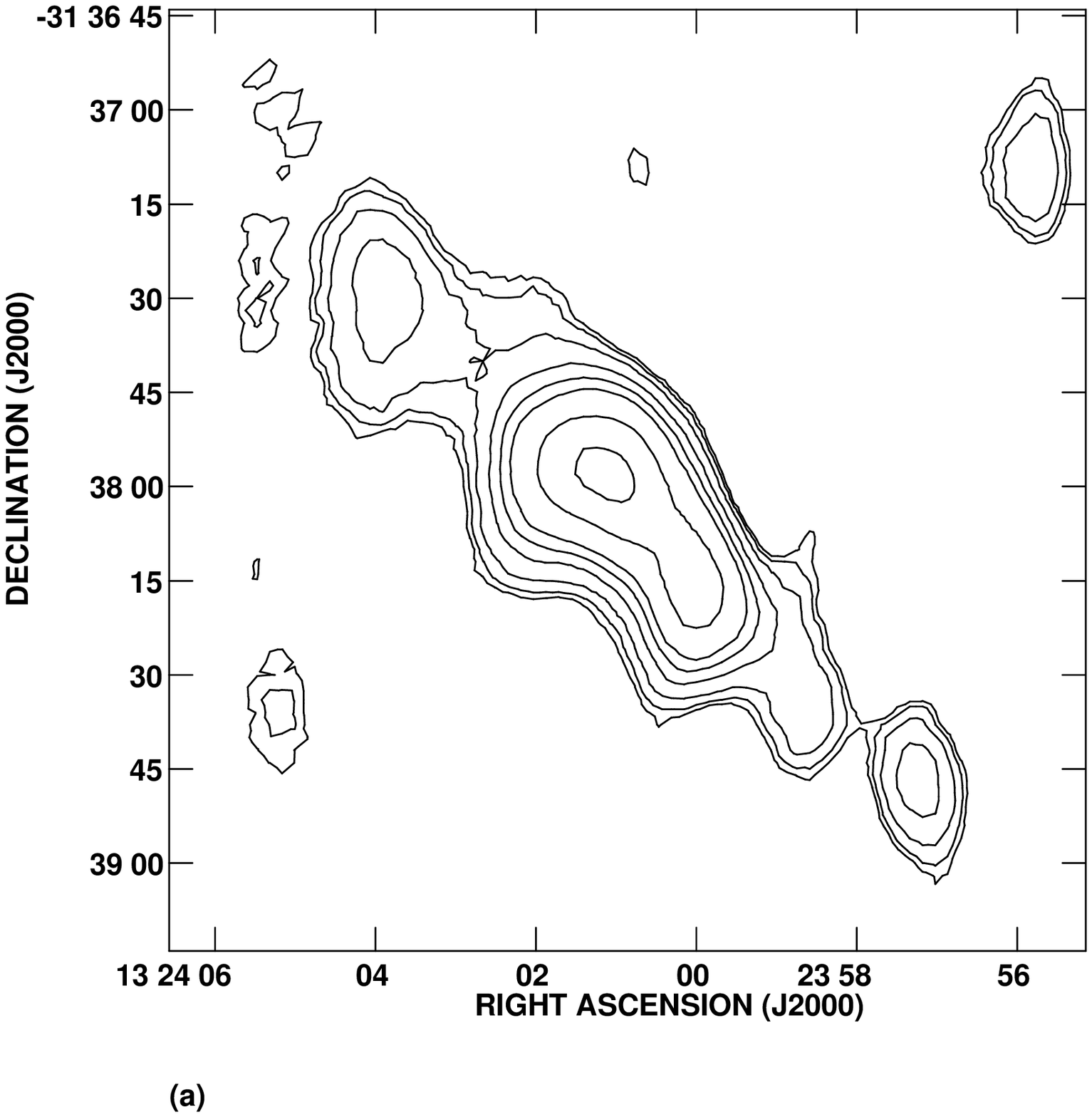}
\epsfysize=8.5cm
\epsfbox{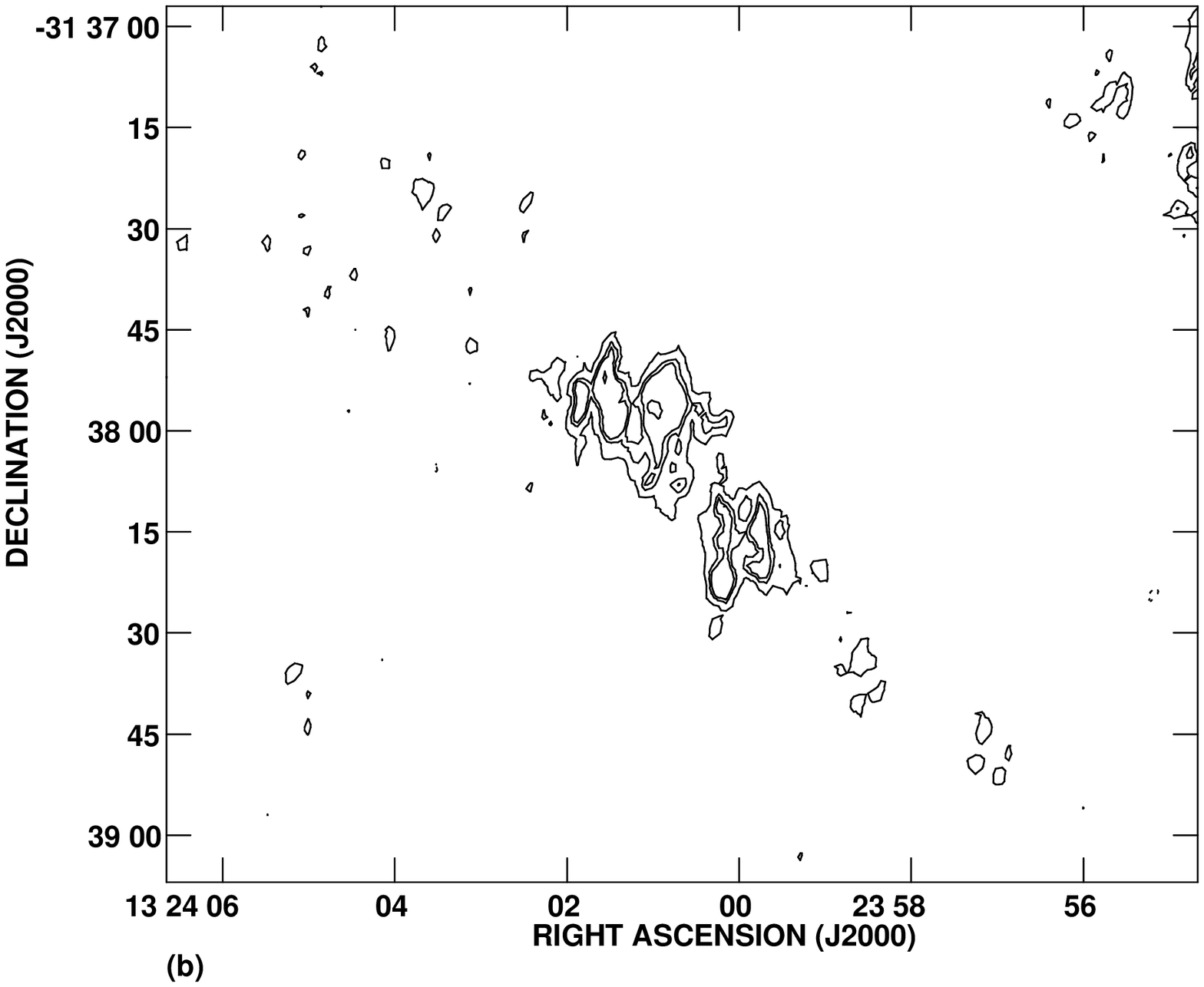}
\caption[]{
(a) Natural weighted 4790 MHz ATCA image.
The resolution is $20^{\prime\prime} \times 10^{\prime\prime}$, 
p.a. $0.42^{\circ}$.
The total flux density is 7.3 mJy. Contour levels
are -0.125, 0.125, 0.15, 0.2, 0.3, 0.4, 0.5, 0.75, 1 mJy/beam.
The rms noise is 0.04 mJy/beam.
\par\noindent
(b) Full resolution 4790 MHz ATCA image.
The resolution is $8.1^{\prime\prime} \times 2.8^{\prime\prime}$, 
p.a. $-5.5^{\circ}$.
The total flux density is 7.0 
mJy. Contour levels
are -0.1, 0.1, 0.15, 0.175, 0.2 mJy/beam.
The rms noise is 0.04 mJy/beam.}
\end{figure}
%
%
%
%
%
%
% FIGURE 7.
\begin{figure}
\epsfysize=8.5cm
\epsfbox{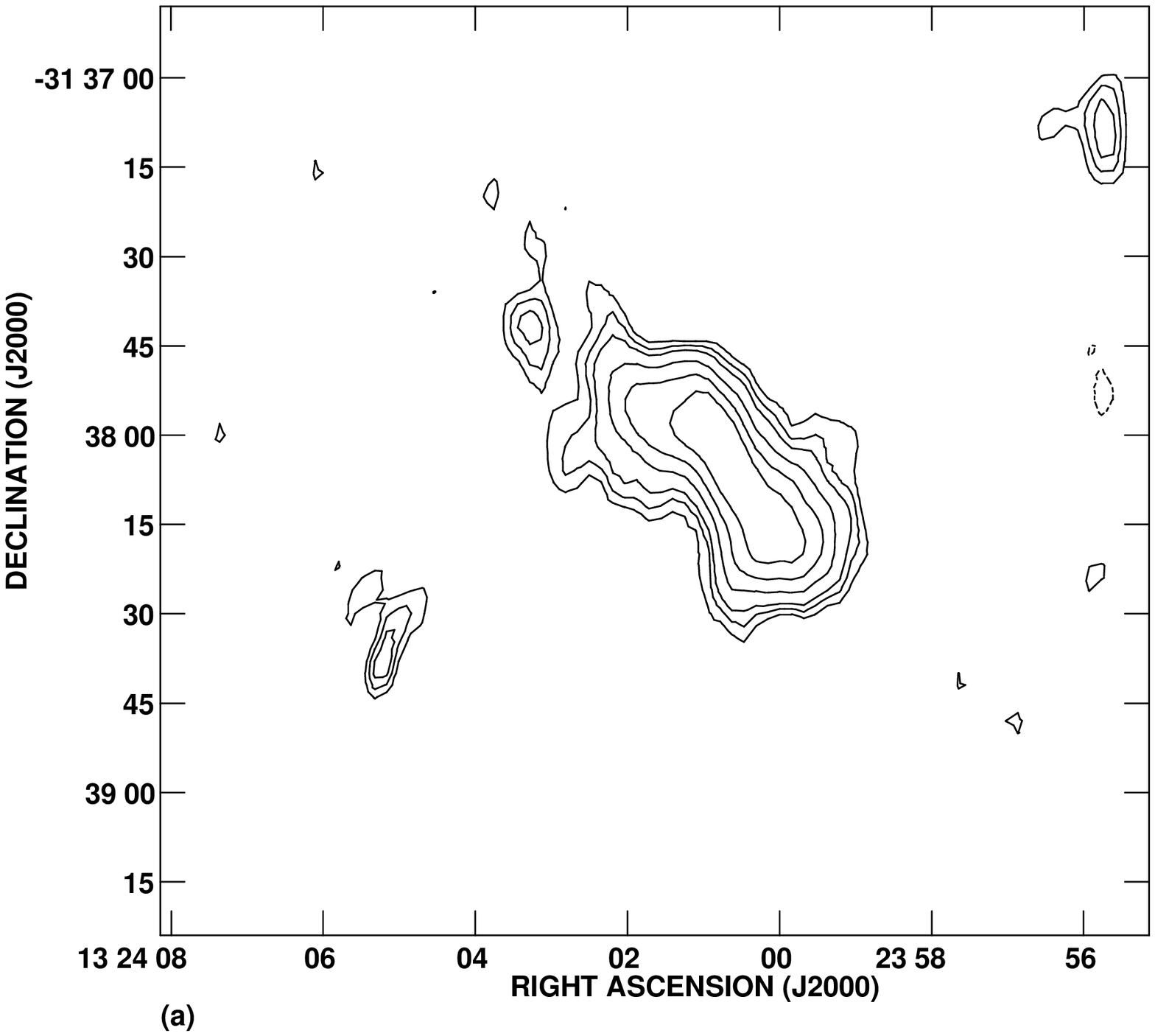}
\epsfysize=8.5cm
\epsfbox{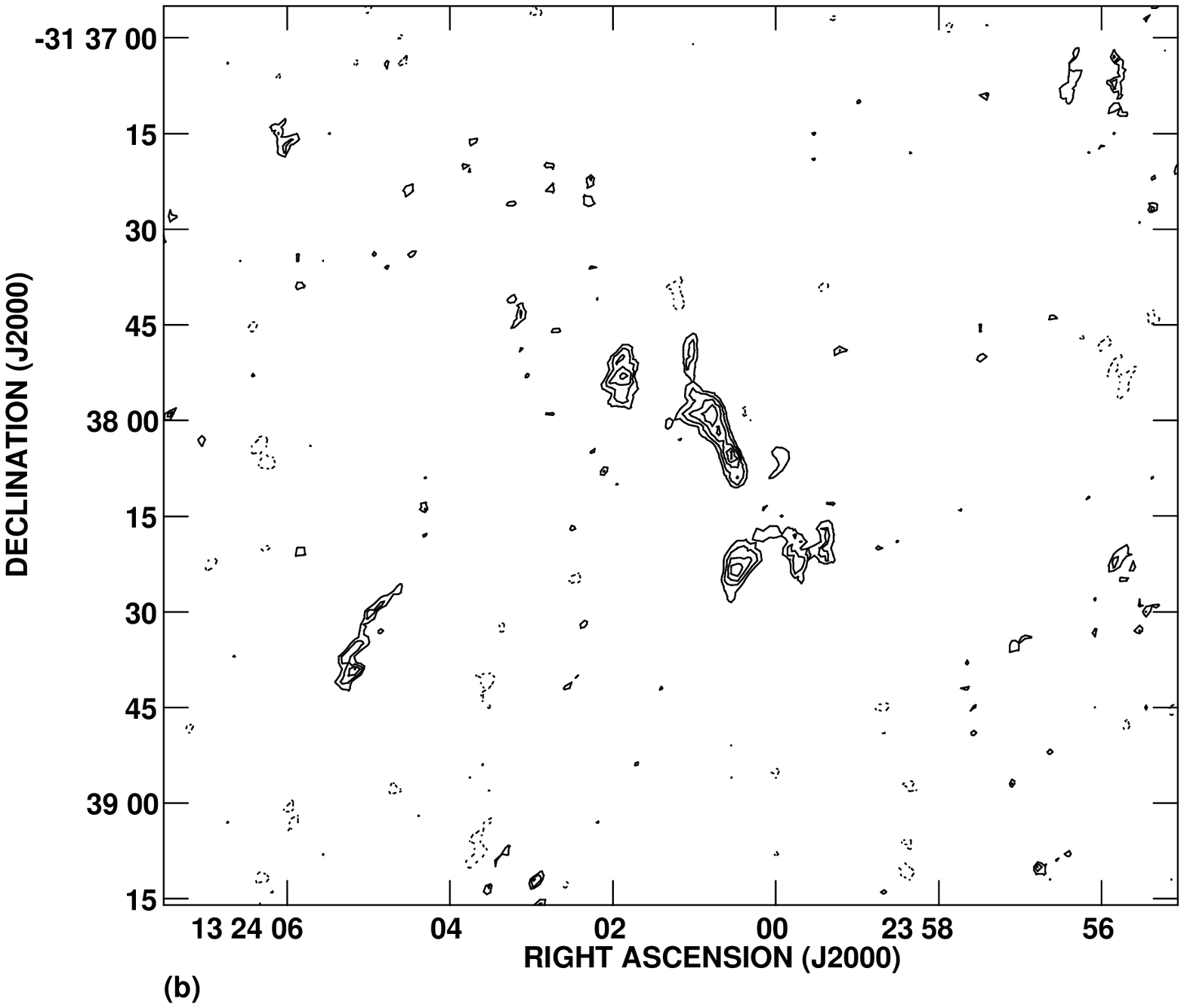}
\caption[]{
(a) Natural weighted 8640 MHz ATCA image.
The resolution is $20^{\prime\prime} \times 10^{\prime\prime}$, 
p.a. $0^{\circ}$.
The total flux density is 2.2 mJy. 
Contour levels
are -0.1, 0.1, 0.125, 0.15, 0.2, 0.25, 0.3 mJy/beam.
The rms noise in the image is 0.04 mJy/beam.

\par\noindent
(b) Full resolution 8640 MHz ATCA image.
The resolution is $9.5^{\prime\prime} \times 3.7^{\prime\prime}$, 
p.a. $-0.53^{\circ}$.
The total flux density in the
present image, computed by integrating over the same area as in Figure 7a,
is 2.1 mJy. Contour levels are -0.1, 0.1, 0.125, 0.15, 0.175 mJy/beam.
The rms noise is 0.04  mJy/beam.}
\end{figure}
\subsection{Polarisation information}

Using the final ATCA data sets at all frequencies we made full resolution
images in the U and Q
Stokes parameters, in order to image the polarised emission. No polarised 
emission above the noise level 
%($\sim 35$ $\mu$Jy at all frequencies)
was detected from the source. 
Integrating the flux density on the polarisation images over the same
area covered by the source in the total intensity ones, we derive
a limit on the polarisation which ranges from 
a few percent (up to $\sim$ 4\%) at 1376 MHz and 4790 MHz to 
$\sim$ 15\% at 8640 MHz.

\vskip 1.0truecm
\noindent
\section{The radio morphology}

Inspection of figures 2 to 7 shows that the appearance of 
J1324$-$3138 changes markedly at different frequencies and resolutions.
The source is barely elongated in the north-east direction at 327 MHz 
and 843 MHz, then the morphology becomes more diffuse and amorphous with 
increasing resolution and frequency. 
Our 327 MHz and 843 MHz images are in very good agreement with the 1.4
GHz NRAO VLA Sky Survey (NVSS) image of the source. 
From the 1376 MHz images we derived a projected linear extension is 
$\sim 82 \times 15$ kpc. From the figures it is clear that the quality of 
the 2382 MHz image is lower than
the other images, and this is essentially due to
the poorer uv-coverage. 
At 2382 MHz we observed J1324$-$3138 only with the 6A array 
configuration, and the lack of short baselines is responsible for the
knotty appearance of the source at this frequency.

\noindent
As is clear from the images and from the flux density measurements
(see figure captions),
when the source is observed with enough resolution
to be resolved into its various components (at frequencies higher than 1376 MHz 
in our set of observations) the total flux 
density is dominated by the extended emission located outside 
the optical galaxy.
The southernmost component, coincident with the galaxy \#5975, is resolved
in all our images, and its emission becomes very faint and barely
visible in our low resolution 8640 MHz map.

\noindent
From our images we classify J1324$-$3138 as a head-tail radio galaxy.
Its total radio power at the canonical frequency of 
1.4 GHz is logP$_{1.4 GHz}$ = 23.05 W Hz$^{-1}$.
The source therefore falls within the class
of FRI (Fanaroff \& Riley 1974), as usually found for head-tail radio
sources. The properties of the core of the radio emission, whose 
power is logP$_{1.4 GHz}$ = 21.75 W Hz$^{-1}$, are presented in detail
in {\it Sect. 4.2}.

\medskip
Some diffuse and faint radio emission, located south of the optical galaxy, 
is visible in some of our images, in particular the 327 MHz VLA image 
(Fig. 2) and in the low 
resolution 1376 MHz and 4790 MHz images (Fig. 4a and Fig. 1 respectively). 
In the 327 MHz image the feature
is elongated in the east-west direction, and possibly is a blending of 
sources containing also radio emission from the cD galaxy at the cluster centre.
Inspection of the 4790 MHz image superimposed on the Digitised Sky Survey,
showed that the radio emission located between the galaxy \#5975
and the cD galaxy at the cluster centre has a faint optical counterpart,
which is also visible in Figure 1.
We therefore consider this emission unrelated to J1324$-$3138 and do not
take it into account in the discussion.

\vskip 1.0truecm
\noindent 
\section{\bf 4. Radio Spectrum}

\subsection{The integrated spectrum}

Using the total flux density measurements reported in Table 2 
we derived the total
spectrum of J1324$-$3138 in the frequency range 327 MHz - 8640 MHz, shown 
in Figure 8. 
The spectrum is steep over the whole frequency range covered by our
observations, but a single power law is a poor fit.
We can fit a single power
law from 843 MHz to 4790 MHz, with a spectral index 
$\alpha_{0.84 GHz}^{4.79 GHz} = 1.3$ (S$\propto \nu^{-\alpha}$), but the
shape of the spectrum steepens at higher frequencies and tends to
flatten below 843 MHz. Given the diffuse morphology of the source, we point 
out that some extended flux could have
been missed at high frequencies (i.e. 4790 MHz and 8640 MHz), and that
the flux density used at these two frequencies should be considered
lower limits.

We have fitted the integrated spectrum of J1324$-$3138, considering
synchrotron radiative losses, in order to derive some information on the 
intrinsic nature
of the source and on the age of the radiating electrons. We have used the
program SYNAGE (Murgia \& Fanti 1996) and have considered three different
possibilities, i.e. {\it (a)} continuous injection of new radiating electrons, 
{\it (b)} constant pitch angle of the radiating electrons 
(Kardashev 1962, Pacholzyck 1970, 
hereinafter KP model); {\it (c)} reisotropisation of the electrons, i.e.
variable pitch angle distribution (Jaffe \& Perola 1973, hereinafter JP model).

\noindent
Our spectrum is very well fitted by both the KP and by the JP models, while 
the continuous injection model does not seem to apply to our
data. Both KP and JP models converge for an initial spectral index
$\alpha_{inj}$ (spectral index of the synchrotron radiation in the
part of the spectrum not affected by the evolution) of the order of
$\alpha_{inj} = 0.9 \pm 0.1$, which implies a power law for the energy 
distribution of the injected electrons of $\delta = 2.8 \pm 0.2$.
The corresponding break frequencies for the two models are $5.1 \pm 1.3$ GHz
(KP) and $10.1 \pm 2.8$ GHz (JP). The fit of the JP model to our
observational data is also shown in Fig. 8.

\noindent
An estimate for the source age $t_{sync}$ was possible after computation of 
the physical parameters within the source, i.e. the magnetic field,
the internal non-thermal pressure and the internal energy density.
We assumed a cylindrical geometry in the source and equipartition conditions,
with a uniform filling factor ($\Phi$ = 1), a ratio K between protons and 
electrons of 1, and the spectral index given by our observations.
We derived P$_{int} = 1.5 \times 10^{-13}$ dyne cm$^{-2}$, 
u$_{min} = 2.5 \times 10^{-13}$ erg cm${-3}$ and B$_{eq}$ = 1.6 $\mu$G
(see also Paper I).
The derived magnetic field  leads to an age estimate of 
$t_{sync} \sim 10^8$ yrs, including inverse Compton losses
and the equivalent field of the 3K radiation 
(H$_{cbr} = 3.25 \times (1+z)^2$ $\mu$G).
%
%
%
% FIGURE 8.
\begin{figure}
\epsfysize=8.5cm
\epsfbox{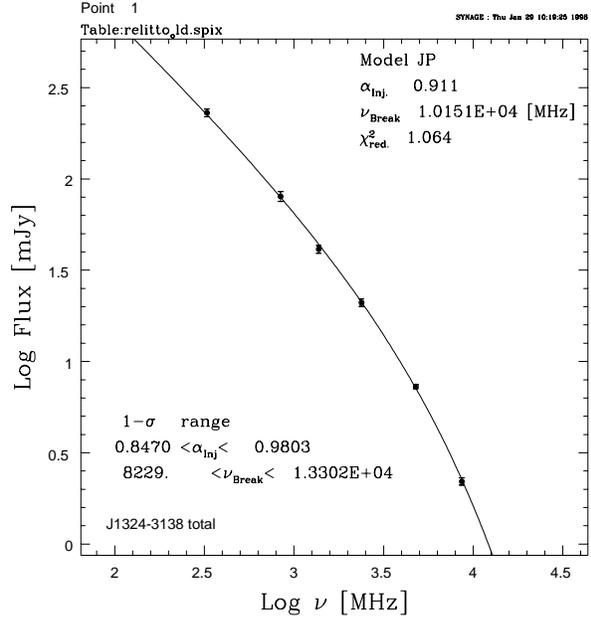}
\caption[]{
Total radio spectrum of J1324$-$3138 obtained with our data
(see Table 2). The continuum line shows the fit of the Jaffe \& Perola
model (see Section 4.1 in the text) to the dataset.}
\end{figure}
%
%
%

%
%------ TABLE 2
%
\begin{table}
\caption[]{ Parameters of the images}
\begin{flushleft}
\begin{tabular}{rcclc}
\hline\noalign{\smallskip}
 Frequency &  HPBW & rms   & S$_{tot}$ & S$_{core}$ \\
   MHz     &$^{\prime\prime}$,$^{\circ}$ & mJy/b & mJy  & mJy \\
\noalign{\smallskip}
\hline\noalign{\smallskip}
  327   & $59 \times45$, 80.5    & 1.9  & 230   & -        \\
  843   & $82 \times43$,  0.0    & 1.4  & 80    & -        \\
 1376   & $10.2 \times5.9$, 0.3  & 0.14 & 41    & 2.05     \\
 2382   & $9.8 \times 6.5$, 0.0  & 0.16 & 21    & 1.74     \\
 4790   & $20 \times 10$, 0.0    & 0.04 & 7.3   & 0.46     \\
 8640   & $20 \times 10$, 0.0    & 0.04 & 2.2   & 0.12     \\
\noalign{\smallskip}
\hline
\end{tabular}
\end{flushleft}
\end{table}
\subsection{Spectrum of the nuclear component}

The spectrum of the component coincident with the optical
galaxy, obtained using the flux density measurements given in Table 2,
is shown in Figure 9. Unfortunately we could not include flux density values
below 1376 MHz, since the resolution of our images
is too low to separate the contribution of the nuclear component from the
rest of the emission.
The spectrum is very steep for frequencies higher than 2382 MHz,
with a value $\alpha_{2.38 GHz}^{8.64 GHz} = 2.1\pm 0.3$, while it
flattens considerably between 1376 and 2382 MHz, where 
$\alpha_{1.38 GHz}^{2.38 GHz} = 0.3\pm0.4$. The poor quality 
of the 2382 MHz map reflects in a high uncertainty in the spectral index
values. We point out here that
the main uncertainty in the computation of the flux density
of the various components in J1324$-$3138 comes from their diffuse appearance.
The overall shape of the spectrum, i.e. flat to normal below 2382 MHz and very
steep above this frequency, suggests that this component is the core of 
the radio emission.

\noindent
As in the case of the integrated spectrum, we have fitted the 
steep part of the spectrum with the program SYNAGE.
Here the JP model seems to fit the observed data best.
We derive $\alpha_{inj}$ in the range 0.85 - 1.0, in agreement with the
value found for the integrated spectrum. The break frequency is lower
here, with $5.2 < \nu_{break} < 7.5$ GHz, with the best fit for 
$\nu_{break} = 7.0$ GHz. The magnetic field computed
under the assumption of equipartition is B$_{eq} = 2.4~ \mu$G, 
u$_{min} = 4.6 \times 10^{-13}$ erg cm$^{-3}$ and 
P$_{min} = 2.8 \times 10^{-13}$ dyne cm$^{-2}$. The corresponding 
estimate for the synchrotron age of the radiating electrons in
this component is $9.7 \times 10^7$ 

%
%
%
% FIGURE 9.
\begin{figure}
\epsfysize=8.5cm
\epsfbox{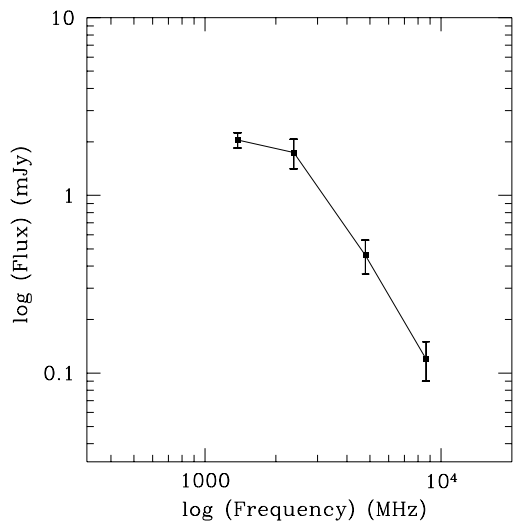}
\caption[]{
Radio spectrum of the core in J1324$-$3138.}
\end{figure}
\subsection{Spectrum along the diffuse tail of radio emission}

The study of the point-to-point spectral index in J1324$-$3138, essential
for studying the evolution of the radio emission along the tail,
is not trivial, given the diffuse morphology of the source and 
its resolution perpendicular to the axis of the radio 
emission in our ATCA images. In order to derive at least the trend of the 
spectral index between 1376 MHz and 2382 MHz along the radio tail, we computed 
$\alpha$ in the two main brightness
peaks located at $\sim 45^{\prime\prime}$ and $\sim 70^{\prime\prime}$ 
from the core. 
There is tentative evidence that the spectral index steepens away from the core,
with $\alpha_{1.38}^{2.38} = 1.3 \pm 0.2$ 
and $1.7 \pm 0.2$ respectively at 
$\sim 45^{\prime\prime}$ and $\sim 70^{\prime\prime}$ .

The smoother morphology of the source in the natural weighted images at 
4790 MHz and 8640 MHz, shown in Figs. 6a and 7a, allows a study 
of the point-to-point spectral index. The two images were made 
with the same gridding and restored with the same beam.
Given the morphology of the 8640 MHz radio emission we could compute the
spectral index starting from a distance of $\sim 40^{\prime\prime}$ from
the core, out to $\sim 100^{\prime\prime}$. The plot of 
$\alpha_{4.79}^{8.64}$ versus distance from the core is given in 
Figure 10.
The spectral index
is steep. From the plot it is clear that $\alpha$ steepens outwards 
along the jet. The local minima in $\alpha$ correspond to the brightness peaks
in the jet, and the values derived in this frequency range are in agreement
with those derived between 1376 MHz and 2382 MHz.
%rom Figs. 7 and 8 we can see that $\alpha_{4790 MHz}^{8640 MHz} = 2.3$ 
%n the core, i.e. is considerably steeper here than along the jet.
%
%
% FIGURE 10.
\begin{figure}
\epsfysize=8.5cm
\epsfbox{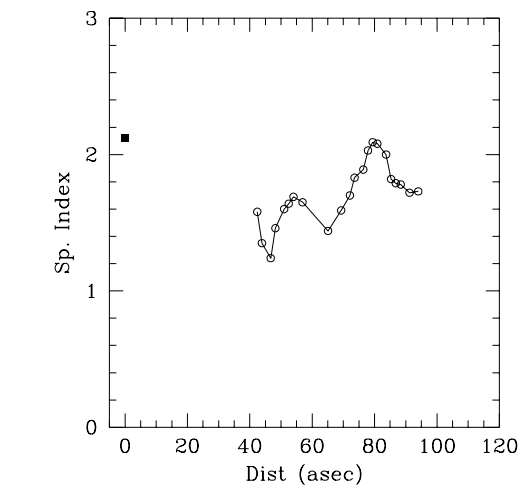}
\caption[]{
Point-to-point spectral index between 4790 MHz and 8460 MHz
along the ridgeline in the tail of emission. The filled square gives
the spectral index of the nuclear component in this frequency range,
for comparison. The two relative peaks in the plot correspond to the
peaks in the brightness emission at 1376 MHz and 2382 MHz.}
\end{figure}
\vskip 1.0truecm
\noindent
\section{Discussion}

The most important observational properties of J1324$-$3138 we should take into
account in interpreting the nature of the source and its implications are:

\noindent
{\it a)} its location in a cluster of galaxies;

\noindent
{\it b)}  its tailed radio morphology, characterised by diffuse emission 
rather than by well defined and/or collimated jets;

\noindent 
{\it c)} a steep total spectral index, with 
$\alpha_{0.84}^{4.79} = 1.3$,
with evidence of further steepening at frequencies above 4790 MHz,
and a steep spectral index in the core, with indication of a flattening around 
1.4 GHz;

\noindent
{\it d)} unpolarised emission down to the noise level at all frequencies.

\subsection{The nature of J1324$-$3138: a relic of a radio galaxy?}

The amorphous appearance of the source, its steep spectrum and the age 
(time since the last acceleration of electrons) derived 
in Section 4, indicate that J1324$-$3138 is an old radio source. 
The total spectral index of J1324$-$3138 in the frequency range covered by our 
observations is similar to that found in the outermost regions
of tailed radio sources in clusters of galaxies (see, for example, what
is found for the extended sources in the peripheral regions of the Coma
cluster [Venturi, Feretti \& Giovannini 1989], and the recent
study carried out on four cluster extended radio
galaxies by Feretti et al. 1998), and to the values derived 
for the few relic sources known (Harris et al. 1993, Feretti \& Giovannini
1996).

\noindent
The presence of the low frequency flattening in the
spectrum of the radio component coincident with the optical galaxy suggests
that it is the nuclear component of the whole radio emission, reinforcing 
the conclusions we drew in Paper I that J1324$-$3138 is a tailed radio galaxy.

\noindent
The values we obtained for the intrinsic physical parameters of the source
are intermediate between what is commonly found along
the tails of head-tail radio sources in clusters of galaxies and in
the few relic radio sources known thus far. For example, typical
values of the internal non-thermal pressure 
$P_{int}$ for a large sample of tailed radio sources in clusters
of galaxies are in the range $0.5 - 5 \times 10^{-12}$ dyne cm$^{-2}$,
with few exceptions (Feretti, Perola \& Fanti 1992).
Values found for relic sources are in the range $2.3 \times 10^{-14}$ - 
$1.5 \times 10^{-13}$ (see Harris et al. 1993 and 
Feretti \& Giovannini 1996, for a recent review). 

The properties of the nuclear component, and of the source as a whole,
led us to the conclusion that no injection of new electrons is taking place at
present, i.e. the nuclear engine has switched off, 
and J1324$-$3138 is a {\it dead} tailed radio galaxy, whose final 
evolution is now dominated by synchrotron losses, possibly in a dense 
confining medium.

\subsection{Location of the source within A3556 and properties
of the intracluster gas}

A3556 is a richness class 0 cluster, with a mean radial velocity
of $<v>=14357$ km/s and a velocity dispersion of $\sigma=643$ km/s
(Bardelli et al. 1998). This cluster is located at the centre of the Shapley 
Concentration supercluster, in a region with clear signs of dynamical activity
(Bardelli et al., 1998).
In particular, A3556 is at the western end of 
a chain of strongly interacting clusters,
which includes also A3558 and A3562 (Bardelli et al. 1994). Its galaxy 
luminosity function shows a plateau for magnitudes brighter than 
$M_{b_J} =-19.8$, 
indicating an excess of bright objects (Paper I). The analysis of the velocity 
histogram in A3556 showed the presence of two significant peaks, 
one corresponding to the main component 
($<v>$=14130 km sec$^{-1}$ and $\sigma=411$ km sec$^{-1}$)
and the other to a group 
($<v>$=15066 km sec$^{-1}$ and $\sigma=222$ km sec$^{-1}$).
The two systems are aligned along the line of sight.
The optical counterpart of J1324$-$3138 is the brightest member of the 
subgroup (it is at rest with respect to the velocity centroid) and its 
optical spectrum (Stein, private communication) is typical of an early type 
galaxy, with a red continuum and no signs of activity, such as 
emission lines (Stein 1996).

The radio source is located at a projected distance of $\sim 2^{\prime}$ 
(corresponding to 0.06 Abell radius) from the dominanant cD galaxy at the 
centre of A3556, but given the dominance of systemic 
peculiar velocities over the Hubble flow, it is impossible to determine
the relative spatial position between A3556 and the subgroup, in particular
whether the systems are in an early stage of merging or the J1324$-$3138 group
is already well inside the main cluster. 
The head-tail appearance of the radio source seems to point toward the latter 
hypothesis, where the morphology arises from the interaction via ram 
pressure with the hot gas of A3556.

Ettori, Fabian \& White (1997), in their extensive study of the X-ray 
properties of the clusters in the Shapley Concentration core, applied 
a deprojection method to a ROSAT image of A3556 in order to derive the 
properties of the hot gas.
By fitting a King model of the form
$$ P_{hg}= P_0 \left[ 1+ \right( r/r_c \left)^2 \right]^{-\alpha} $$
to the deprojected pressure profile, Ettori (private communication)
obtains
$P_0=7.50 \times 10^{-12}$ dyne cm$^{-2}$, $\alpha=0.629$ and $r_c=0.177$.
The equilibrium between the external hot gas pressure and the 
internal equipartition pressure in J1324$-$3138, i.e.
$P_{hg}=P_{int}$, is found by extrapolating the profile out to a 
distance of 
$\sim 4$ Mpc for the extended component of J1324$-$3138 and to $\sim 2.5$ 
Mpc for its core. We point out that these distances are 
significantly larger than the dimension of the X-ray emitting region of A3556
($\sim 0.7$ Mpc) and are larger than the distance between the cores of A3556 
and A3558 ($\sim 2$ Mpc). We conclude therefore that
$P_{hg} > P_{int}$ over the full extent of the radio source and at any 
plausible distance from the centre of A3556.

\noindent
Moreover, we calculated the ram pressure assuming a gas temperature of $KT=
1.6$ keV derived from the velocity dispersion, and a galaxy peculiar
velocity given by the difference between the mean velocity of the A3556 
main component and the 
measured redshift of 
J1324$-$3138 ($\sim 900$ km/s), finding that $P_{ram}>>P_{int}$ at all
distances from the cluster centre.

The condition of pressure equilibrium between the radio emission and the
surrounding medium seems to hold in some cases,
for example in the wide-angle tail
in A2717 (Liang et al. 1997). However, from the literature we know
that non-equilibrium between the external gas thermal pressure and the internal
non-thermal pressure of the radio emission is not uncommon. In particular,
the radio emission is often overpressured, i.e. P$_{hg} >$ P$_{int}$
up to about two orders of magnitude (Feretti, Perola \& Fanti 1992), and
deviations from minimum energy arguments could be expected and should be
taken into account. The unbalance between $P_{int}$ in J1324$-$3138 and  
$P_{hg}$ at its projected distance from the centre of A3556, amounts to a 
factor of $\sim 50$, fully in agreement with the result of Feretti, Perola 
\& Fanti (1992).

\medskip
Further evidence that the source is located in a dense medium and that the
radio emission is possibly seen through the cluster magnetic field
comes from two observational properties, the source age and the 
unpolarised radio emission.
The age estimate we derived in {\it Sect. 4.1} is amongst the
largest found in radio sources. Estimates for the very few relic radio sources
known and for cluster radio halos range around a few times 10$^7$ yrs 
(see for example the case of 1257+275 located in the peripheral region of the 
Coma cluster, Giovannini, Feretti \& Stanghellini 1991). In the case
of the radio galaxy B2 0924+30, considered ``a prototypical genuine relic
of a dead radio galaxy'' (Klein et al. 1996) the ages derived are in the
range 7 - 8 $\times 10^7$ yrs.
For J1324$-$3138 strong confinement by the intracluster medium
may have reduced the losses due to adiabatic expansion, thus allowing a
longer lifetime for the source.

The lack of polarisation from the source may also give information
on the intracluster gas. We note that the upper limit
to the polarisation percentage at 8640 MHz is high, due to the very low
total flux of the source, so from our data we cannot distinguish between
low polarisation at high frequencies combined with depolarisation at high
frequency, or no polarisation at all.
In both cases the observational data could be explained either {\it a)} with
the presence of a large quantity of thermal plasma within the radio source
(Burn 1966), or {\it b)} with an external foreground screen,
characterised by a random magnetic field (Tribble 1991), or both.
We point out that extended sources in clusters of galaxies are 
usually polarised, and that the polarisation percentage 
may go up to $\gtrsim 50\%$ in the outermost regions (Feretti et al. 1998).
Relic radio sources in clusters of galaxies
are polarised too (see for example the relic radio source 1253+275 in
the Coma cluster, Giovannini, Feretti \& Stanghellini 1991). Halo radio sources
at the centre of galaxy clusters, on the other hand, are not polarised.

\noindent
{\it a)} The large age estimated for the source,
and its morphology, suggest that J1324$-$3138 may have had entrained thermal
gas from the ambient medium, and be subject to internal Faraday effects.

\noindent
{\it b)} The existence of cluster magnetic fields is now
an established result. Direct observational evidence comes from the few known
cluster halo sources, such as for example Coma-C in the Coma cluster
(Kim et al. 1989, Giovannini et al. 1993) and A2319 (Feretti, Giovannini
\& B\"ohringer 1997); indirect evidence is given by the excess rotation
measures observed in a number of central radio sources in clusters of galaxies
(Ge et al. 1994, Feretti et al. 1995).
The location of J1324$-$3138 beyond the main cluster concentration,
coupled with the X-ray emission associated with the cluster, do indeed
support the idea that the radio emission is seen through the intracluster gas.
For comparison, we find that the radio source J1324$-$3140, 
associated with the dominant cD galaxy at the centre of the main concentration 
in A3556, is 12\% and 18\% polarised at 4790 MHz and 8640 MHz respectively.

\subsection{ Possible links between present evolutionary stage of 
J1324$-$3138 and the merging between A3556 and its subgroups}

The age derived for J1324$-$3138 is high for a radio source, but it is 
significantly shorter than the typical timescale for
concluding  merging processes in clusters of galaxies, estimated
to be $\sim 10^9$ yrs (Roettinger et al. 1993). The velocity 
distribution of the galaxies in A3556 shows that 
J1324$-$3138 belongs to a subgroup located beyond (in velocity space)
the core. However, from the
spectroscopic information it is impossible to infer whether the subgroup is
still infalling into the main concentration of A3556 or if it has already 
crossed the centre. 
If we assume that cluster merging strongly influences the radio activity in
galaxies, then our study at radio wavelengths suggests two possible 
scenarios.

\noindent
{\it 1.} The galaxy associated with J1324$-$3138, a pre-existing 
radio source, is falling towards the centre
of A3556. This implies that the ram pressure stripped the tail, while the 
external thermal gas pressure confined the electrons, preventing losses from 
adiabatic expansion. The estimates given here for the
lifetime of the radio galaxy and for the linear distance of the galaxy with
respect to the core of A3556 agree with each other and with the velocity
difference of the galaxy.
In this frame it should be explained how infalling (through the imbalance
$P_{ram} >> P_{int}$) could switch off a radio active nucleus.
Furthermore, this possibility implies that 
the radio source is located between us and A3556, and under the hypothesis that 
the polarisation properties of the source are due to an external
medium, the requirement of a foreground screen for the depolarization of the 
emission is challenged.

\noindent
{\it 2.} A transit close to the core of A3556 has triggered the radio emission
(Reid et al., 1998), that ceased $\sim 10^8$ years ago. 
A in the former case, the ram pressure has the central role
of stripping the tail. Furthermore, after the merging event between the core
of A3556 and the subgroup hosting J1324$-$3138, the source is now 
beyond the core of A3556 along the line of sight and it is seen through the 
intracluster gas of A3556 itself.

\noindent
With the current data, both at radio and optical wavelengths, it is difficult 
to discriminate between the two possibilities proposed here. The second 
scenario would better fit with the idea that the global process of
merging of clusters and groups of galaxies enhances the probability
that a galaxy becomes a radio source. We recall here that the core of 
A3556 has unusual properties at optical and radio wavelengths. Its optical
luminosity function is quite different from that of the other clusters
in the A3558 complex, presenting a plateau at bright magnitudes, with 
all bright galaxies in the plateau are radio loud (Paper I).
Moreover, the distributions of bright and faint galaxies as a function of 
distance from the cluster centre differ considerably (Bardelli et al. 1998).
\vskip 1.0truecm
\noindent
\section{Conclusions}

We have presented observations of the radio galaxy J1324$-$3138, located 
in the central region of the Abell cluster A3556, over a wide range of
frequencies and resolutions. We can briefly summarise our results as follows:

\noindent
{\it (a)} J1324$-$3138 is an example of a {\it remnant} of a radio galaxy, i.e.
a source in which the engine of the radio emission has switched off. 
The evolution of the radio emission is presently dominated by synchrotron 
losses.

\noindent
{\it (b)} The radio source is not in pressure equilibrium with the intracluster
gas. In particular it is underpressured.

\noindent
{\it (c)} We suggest that the lack of polarisation in the source 
is due to the presence of an intervening Faraday screen, i.e. a cluster 
scale magnetised medium, as it is now often observed in clusters of galaxies,
which depolarises the radio emission.

\noindent
{\it (d)} Under the hypotesis that cluster mergers influence the
radio emission of a galaxy, the properties of J1324$-$3138, coupled
with the peculiarities of A3556 at radio and optical wavelengths
(Paper I and Bardelli et al. 1998), suggest that
the core of A3556 and the subgroup hosting J1324$-$3138 have already
undergone a major merging event.

\vskip 1.0truecm
\noindent
{\bf Acknowledgments}

We wish to thank D. Dallacasa for his suggestions and discussion
while this work was carried out, and R. Fanti for careful reading
of the manuscript.
We are grateful to S. Ettori for providing
unpublished results, and to P. Stein for giving us the spectrum of 
J1324$-$3138. T.V. acknowledges the receipt of two grants 
from CNR/CSIRO (Prot. n. 119816 and Prot. n. 088864).

The MOST is operated by the University of Sydney,
with support from the Australian Research
Council. 
The Australia Telescope Compact Array is operated by the CSIRO Australia
Telescope National Facility. The National Radio Astronomy Observatory
(NRAO) is operated by Associated Universities, Inc., under contract with the
National Science Foundation.
This research has made use of the NASA/IPAC Extragalactic Database (NED),
which is operated by the Jet Propulsion Laboratory, Caltech, under contract
with the National Aeronautics and Space Administration.

{}
\end{document}